\documentclass[letterpaper,twocolumn,10pt]{article}
\usepackage{usenix}

\usepackage{tikz}

\usepackage{times}
\usepackage{graphicx}
\usepackage{booktabs}
\usepackage{amsmath}
\usepackage{amssymb}
\usepackage[table]{xcolor}
\usepackage{listings}
\usepackage{enumitem}
\usepackage{adjustbox}
\usepackage{bm}
\usepackage[most]{tcolorbox}

\setlist[itemize]{leftmargin=*}
\setlist[enumerate]{leftmargin=*}
\usetikzlibrary{backgrounds}

\lstdefinelanguage{Solidity}{
  morekeywords={contract,function,mapping,struct,if,revert,return,external,
    internal,view,payable,memory,calldata,storage,public,private,uint256,
    uint32,uint64,bytes32,bytes,address,bool},
  morecomment=[l]{//},
  morestring=[b]",
}
\newtcolorbox{findingbox}[1][]{%
  colback=black!5,
  colframe=black,
  boxrule=1pt,
  arc=0pt,
  auto outer arc,
  boxsep=0pt,                  
  left=3mm, right=3mm,         
  top=2mm, bottom=2mm,         
  width=\linewidth-3pt,  
  fontupper=\footnotesize,
  #1
}

\newcommand{\toolname}{IntentFuzz}

\begin{document}

\date{}

\title{\Large \bf IntentFuzz: A Protocol-Aware Fuzzer for Automated Invariant\\Violation Detection in Intent-Based Cross-Chain Bridges}


\author{
{\rm André Augusto}\\
INESC-ID \& IST, University of Lisbon
\and
{\rm Christof Ferreira Torres}\\
INESC-ID \& IST, University of Lisbon
\and
{\rm André Vasconcelos}\\
INESC-ID \& IST, University of Lisbon
\and
{\rm Miguel Correia}\\
INESC-ID \& IST, University of Lisbon
} 

\maketitle


\begin{abstract}
Cross-chain bridges move value between blockchains. Intent-based bridges are a variant where a solver fulfills a user's declared outcome and an off-chain settlement layer later reconciles the fill against the deposit. Existing smart-contract fuzzers and static analyzers only flag known-bad code patterns or require protocol-specific hand-written assertions. This work formalizes a taxonomy separating invariant violations, safety properties a contract must enforce locally, from settlement exposures legitimately delegated to the off-chain settlement layer, and proposes \textbf{\toolname{}: a protocol-aware fuzzer} that recovers a bridge's intent structure and deposit/fill function roles directly from unannotated Solidity source, then synthesizes multi-step fuzz sequences using an LLM-based fallback to help build call arguments. \toolname{} recovers the correct intent structure in 9/9 benchmark protocols and classifies deposit and fill functions with 100\% recall and 82\% combined precision; across a corpus of 77 manually labeled contracts, it reaches 79.5\% bridge-classification precision and 97.2\% recall, and among confirmed bridges, struct selection reaches 88.6\% precision and recall while deposit and fill classification each reach 100\% recall. On 23 planted-bug mutants, \toolname{} attains 100\% recall and 100\% precision, executing 273 templates (507 transactions in a median of 14ms per template). Across 24 real-world deployments, \textbf{it confirms 17 genuine invariant violations} under heuristic-only input generation, \textbf{rising to 22 with its LLM-assisted tier enabled}, spanning eight vulnerable GitHub repositories, each finding reproducible against public, deployed bytecode.
\end{abstract}

\section{Introduction}
\label{sec:introduction}

Intent-based bridges are cross-chain protocols where users specify the outcome they want, and third-party \emph{solvers} compete to fulfill that \emph{intent} by executing the necessary actions across chains: for example, a user on Ethereum who wants 1,000 USDC delivered to an address on Arbitrum locks 1,000 USDC in a \emph{deposit} describing that outcome, and a solver fronts the funds on Arbitrum before later claiming reimbursement from the deposit (Section~\ref{sec:background} details this deposit/solver/fill/settlement flow)~\cite{lifi_intents}. This solver-driven design changes the security model compared to normal bridges~\cite{chitra2024analysis}. Cross-chain bridges collectively secure billions of dollars in locked value and have been the target of some of the largest exploits in decentralized finance, with total losses exceeding \$3.2 billion~\cite{lee2023sok,zhang2024sok,augusto2024sok,10.1007/978-3-032-32575-4_16}. This makes the correctness of any new architecture these protocols adopt a high-stakes question. ERC-7683~\cite{erc7683}, recently proposed as a standard interface for intent-based bridges, reflects this architecture's growing adoption, a trend this work's own corpus of GitHub repositories independently corroborates (Section~\ref{sec:eval-corpus}).

Existing smart contract fuzzers (rooted in the tradition of randomized testing~\cite{Miller:90}) are effective at finding bytecode or source code defects inside a single contract execution~\cite{wu2024we}, but they fall short for intent-based bridges. Intent-based bridges are instead defined by cross-chain semantics, off-chain solver behavior, and multi-blockchain flows. In this setting, the space of possible failures extends beyond malformed inputs or malicious instruction sequences: it includes replay across chains, partial or delayed settlement, and other protocol-specific correctness properties that only emerge once multiple chains and an off-chain actor are considered together. Traditional fuzzers typically treat a test as a transaction or short on-chain sequence, and their assessments are usually tied to coverage, assertions, or generic vulnerability patterns~\cite{jiang2018contractfuzzer,torres2021confuzzius}. As a result, they cannot reliably infer the intended bridge semantics for any of these failure modes.

The security properties of intent-based bridges are qualitatively different from those of conventional smart contracts. A bridge is correct if and only if the assets received by the beneficiary on the destination chain match the intent declared on the source chain: same token, same amount, same recipient, within the declared deadline, and non-replayable. We propose a two-tier finding taxonomy. \emph{Invariant violations} occur where the on-chain contract fails to enforce a safety property regardless of solver behavior (e.g., the same intent processed twice or after the deadline has passed). \emph{Settlement exposures}, by contrast, occur where the on-chain contract never checks a correctness property at all, such as whether the recipient, token, or amount a fill actually delivers matches what the deposit declared, because the protocol relies entirely on its off-chain settlement layer to catch a mismatch after the fact; if that settlement layer is itself compromised or buggy, the mismatch goes uncaught.

We present \toolname{}, a protocol-aware fuzzer for intent-based bridges. Given an unannotated Solidity implementation of a bridge, \toolname{} uses static analysis to extract the data structure (i.e., Solidity \texttt{struct}) that represents an intent, identify its correctness properties based on the fields present, and classify protocol functions by semantic role (e.g., deposit, fulfillment). It then constructs multi-step fuzz templates that pair source-chain deposits with destination-chain fulfillments. To handle the diversity of intent protocols, \toolname{} infers protocol structure and safety properties automatically from source code, supporting both single-contract and cross-contract deployments. It then generates one fuzz suite per deposit-fulfillment function pair identified, so that all protocol-level entry points are covered. The evaluation of \toolname{} is based on 1) 4 manually written reference contracts with no planted vulnerabilities (\emph{secure baselines}), each paired with a set of mutants in which exactly one invariant check has been deliberately removed, exercising every dynamically-tested invariant in the taxonomy; 2) a corpus-scale analysis of intent-based bridge contracts found in the wild via a GitHub crawler; and 3) 24 real-world protocol/chain deployments reached through forked mainnet state. 

\noindent
\textbf{Contributions.} This paper makes the following contributions:

\begin{itemize}
    \item \textbf{An invariant violation taxonomy for intent-based bridges.} We formalize the invariant-violation/settlement-exposure distinction introduced above into a taxonomy of six invariants that intent-based bridge contracts must enforce on-chain, together with the set of properties that are legitimately delegated to the settlement layer (Section~\ref{sec:invariant-taxonomy}). 

    \item \textbf{Automated recovery of intent structure from unannotated source.} \toolname{} identifies a protocol's intent struct and classifies the semantic role of each field, and separately classifies which functions deposit and which fulfill intents, using only static analysis of Solidity source with no manual annotation. It also flags fill functions gated on a single stored authority as a centralization risk, without requiring code execution. On a benchmark of 9 protocols (5 real-world, 4 synthetic), this recovers the correct struct in 100\% of cases (9/9) and classifies deposit and fill functions with 100\% recall and 82\% combined precision. At corpus scale, across 77 manually labeled contracts, bridge classification reaches 79.5\% precision and 97.2\% recall, struct selection reaches 88.6\% precision and recall, and deposit and fill function classification each reach 100\% recall.

    \item \textbf{Automated multi-step fuzz plan synthesis.} \toolname{} synthesizes multi-step fuzzing sequences directly from the intent structure and function roles recovered by the extraction step described above. This includes a cross-step binding mechanism that lets later steps reference values produced by earlier ones, and a dual-layer tamper mechanism that independently perturbs an ERC-7683 envelope struct and its encoded inner payload to test whether both layers are validated.

    \item \textbf{A three-tier input-generation hierarchy with an LLM-powered feedback loop.} \toolname{} generates call arguments through a session-intent anchor, followed by type-aware boundary sampling, followed by a large language model (LLM)-assisted recovery step. This step is invoked only when a deposit or fill call fails unexpectedly, and its result is cached per function. In an ablation against real-world deployed contracts, this recovery step raises step-0 execution success from 167/307 (54.4\%) to 261/307 (85.0\%) and surfaces 5 additional genuine invariant violations. It acts as a cold-start fallback for protocol-specific input requirements that the lower tiers do not anticipate.
\end{itemize}

Section~\ref{sec:background} covers background; Sections~\ref{sec:invariant-taxonomy}--\ref{sec:evaluation} present the taxonomy, \toolname{}'s design, and its evaluation; Sections~\ref{sec:discussion}--\ref{sec:conclusion} discuss limitations, related work, and conclude.

\section{Background}
\label{sec:background}

In this section, we present the model behind intent-based bridges and the ERC-7683 standard.

\subsection{The Intent-Bridge Model}

The flow of an intent-based bridge is illustrated in Figure~\ref{fig:intent-bridge-model}.

\usetikzlibrary{positioning, arrows.meta, calc, fit, backgrounds}
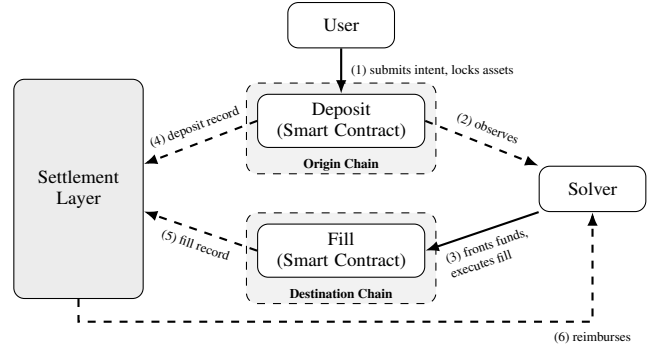
\begin{figure}[t]
\centering
\begin{tikzpicture}[
  every node/.style={font=\scriptsize},
  actor/.style={draw, rounded corners, align=center, minimum height=0.6cm,
                text width=1.2cm, inner sep=3pt, fill=white},
  box/.style={draw, rounded corners, align=center, minimum height=0.65cm,
              text width=2cm, inner sep=3pt, fill=white},
  chain/.style={draw, dashed, rounded corners, inner sep=4pt, minimum height=1.2cm},
  settle/.style={draw, rounded corners, align=center, minimum height=2.9cm,
                 text width=1.5cm, inner sep=3pt, fill=gray!15},
  arr/.style={-{Latex[length=1.8mm,width=1.4mm]}, thick},
  darr/.style={-{Latex[length=1.8mm,width=1.4mm]}, thick, dashed}
]

\node[actor] (user) {User};
\node[box, below=6mm of user] (deposit) {Deposit \\(Smart Contract)};
\node[box, below=10mm of deposit] (fill) {Fill \\(Smart Contract)};
\node[actor, right=15mm of deposit, yshift=-9mm] (solver) {Solver};
\node[settle, left=15mm of deposit, yshift=-9mm] (settlement)
  {Settlement\\Layer};

\draw[arr]  (user) -- node[right, font=\tiny, align=left]
  {(1) submits intent, locks assets} (deposit);
\draw[darr] (deposit.east) -- node[above, sloped, font=\tiny]
  {(2) observes} (solver.north west);
\draw[arr]  (solver.south west) -- node[below, sloped, font=\tiny, align=left]
  {(3) fronts funds,\\executes fill} (fill.east);
\draw[darr] (deposit.west) -- node[above, sloped, font=\tiny]
  {(4) deposit record} (settlement);
\draw[darr] (fill.west) -- node[below, sloped, font=\tiny]
  {(5) fill record} (settlement);
\draw[darr] (settlement.south) -- ++(0,-3mm) -| (solver.south)
  node[pos=0.5, below, font=\tiny]{(6) reimburses};

\begin{scope}[on background layer]
\node[chain, fit=(deposit), yshift=-1mm, fill=gray!10] (originbox) {};
\node[chain, fit=(fill), yshift=-1mm, fill=gray!10] (destbox) {};
\end{scope}

\node[below=0.2mm of deposit, font=\tiny] {\textbf{Origin Chain}};
\node[below=0.2mm of fill, font=\tiny] {\textbf{Destination Chain}};

\end{tikzpicture}
\caption{The flow of information in the intent-bridge model}
\label{fig:intent-bridge-model}
\end{figure}

A user creates an intent (1) by submitting a \emph{deposit} to a smart contract on the chain where the request originates, the \emph{origin chain}. The deposit records the intent's parameters and locks the assets that back it. (2) A set of entities called \emph{solvers} observe the deposit and compete to (3) fulfill the intent by executing the necessary actions on the \emph{destination chain}. The fulfillment is also called just \emph{fill} and requires the solver to front its own funds to execute the requested cross-chain transfer. Because the solver fronts its own funds, it must later be reimbursed (6) from the deposit locked on the origin chain. Reconciling the two chains and authorizing this reimbursement is the role of a \emph{settlement layer}, a protocol-specific off-chain-coordinated mechanism that compares the fill against the deposit (4,5) it claims to satisfy. This verification is necessarily retrospective: the settlement layer observes a fill only after it has already been submitted on the destination chain, and it has no opportunity to review or block a fill before execution. Its function is to adjudicate completed cross-chain activity, not to gate it in advance.

\subsection{The ERC-7683 Standard}

Figure~\ref{fig:erc7683-envelope} illustrates the ERC-7683 standard~\cite{erc7683}, which formalizes the interface of intent-based bridges. ERC-7683 standardizes the origin- and destination-chain interfaces of the intent-based bridge model described above. This lets solvers and settlement infrastructure built for one protocol interoperate with any other protocol that implements the standard, rather than requiring bespoke integration per deployment. The standard defines two entry points for deposits on the origin chain, \texttt{openFor} and \texttt{open}, which accept a \texttt{GaslessCrossChainOrder} and an \texttt{OnchainCrossChainOrder}, respectively. Both envelopes carry a fill deadline (\texttt{fillDeadline}), a type identifier for the inner payload, and the inner payload itself, but they differ in that \texttt{GaslessCrossChainOrder} carries additional fields, including a nonce, an origin-chain identifier, and a separate open deadline (\texttt{openDeadline}) bounding when the order may be opened at all. The destination-chain fill function receives the same encoding back as \texttt{originData}, linking the source- and destination-chain representations of one intent.

The fields that ERC-7683 does not standardize are not part of the envelope itself (e.g., the assets being exchanged, amounts, destination-chain recipients, and any other parameters specific to a given bridge implementation). Instead, the envelope carries them indirectly inside \texttt{orderData}, an opaque, Application Binary Interface (ABI)-encoded \texttt{bytes} field shared by both envelope variants. ERC-7683 specifies only that \texttt{orderData} exists and that its contents are ABI-encoded; it leaves the internal layout of this inner payload entirely to each protocol's own implementation. Consequently, the envelope's directly readable fields are the only ones a generic, protocol-agnostic solver/observer can interpret without further knowledge, decoding \texttt{orderData} requires knowing which protocol issued the order. The fact  that \toolname{} can recover the structure of this inner payload from unannotated source code is a key enabler for its protocol-agnostic fuzzing capabilities.

\usetikzlibrary{positioning, arrows.meta, calc, fit}
\begin{figure}[t]
\centering
\begin{tikzpicture}[
  every node/.style={font=\scriptsize},
  sig/.style={align=left, text width=8.5cm, font=\scriptsize\ttfamily, inner sep=0pt},
  envelope/.style={draw, rounded corners, align=left, text width=3.6cm,
                   font=\scriptsize\ttfamily, inner sep=5pt, fill=white},
  bytesbox/.style={draw, rounded corners, align=center, text width=3.6cm,
                   minimum height=1cm, inner sep=8pt, fill=gray!10},
  arr/.style={-{Latex[length=2mm,width=1.6mm]}, thick}
]

\node[envelope] (gasless) {%
\textbf{struct} GaslessCrossChainOrder \{\\
\ \ address originSettler;\\
\ \ address user;\\
\ \ uint256 nonce;\\
\ \ uint256 originChainId;\\
\ \ uint32 openDeadline;\\
\ \ \textbf{uint32 fillDeadline};\\
\ \ \textbf{bytes32 orderDataType};\\
\ \ \textbf{bytes {\setlength{\fboxsep}{1.5pt}\colorbox{green!20}{\texttt{orderData}}}};\\
\}%
};

\node[envelope, anchor=north west] (onchain) at ($(gasless.north east)+(3mm,0)$) {%
\textbf{struct} OnchainCrossChainOrder \{\\
\ \ \textbf{uint32 fillDeadline};\\
\ \ \textbf{bytes32 orderDataType};\\
\ \ \textbf{bytes {\setlength{\fboxsep}{1.5pt}\colorbox{green!20}{\texttt{orderData}}}};\\
\}%
};

\coordinate (topmid) at ($(gasless.north)!0.5!(onchain.north)$);

\node[sig, above=2mm of topmid] (sigs) {%
  \textbf{function} open(OnchainCrossChainOrder order) external;\\[3pt]
  \textbf{function} openFor(GaslessCrossChainOrder order, bytes signature,
  \ \ bytes originFillerData) external;
};

\coordinate (mid) at ($(gasless.south)!0.5!(onchain.south)$);

\node[bytesbox, draw=none, anchor=north, below=3mm of onchain] (bytes) {%
  {\setlength{\fboxsep}{1.5pt}\colorbox{green!20}{\texttt{orderData}}} is an opaque \texttt{bytes} field with protocol-specific layout%
};

\node[sig, below=20mm of mid] (fill) {%
  \textbf{function} fill(orderId, originData, fillerData) external;%
};

\coordinate (chainliney) at ($(fill.north)+(0,6mm)$);
\coordinate (chainlinel) at (fill.west |- chainliney);
\coordinate (chainliner) at (fill.east |- chainliney);
\draw[thick] (chainlinel) -- (chainliner);
\node[font=\footnotesize, anchor=south east] at (chainliner) {Origin Chain Contract Interface};
\node[font=\footnotesize, anchor=north east, yshift=-0.2mm] at (chainliner) {Destination Chain Contract Interface};

\end{tikzpicture}
\caption{ERC-7683's function signatures and envelope structs.}
\label{fig:erc7683-envelope}
\end{figure}
\section{Invariant Taxonomy}
\label{sec:invariant-taxonomy}

Because the settlement layer's verification is necessarily retrospective (Section~\ref{sec:background}), this timing constraint splits the properties an intent-based bridge deployment must satisfy into two classes, separated by a single test: is the property \emph{locally decidable} (i.e., checkable using only state already resident on the chain that executes the deposit or fill call, without reference to another chain's state)?
Properties that pass this test can be enforced unilaterally by the origin- or destination-chain contract, at the moment the call executes. If a contract fails to enforce such a property, it is broken, regardless of how the solver behaves or what the settlement layer later concludes. We call a failure to enforce such a property an \emph{invariant violation}.
Properties that fail this test, whether the destination-chain recipient, output token, and output amount actually match what the origin-chain deposit specified, are not locally decidable at all: at the moment of the call, the contract has no access to the other chain's state, so no on-chain check at either endpoint can resolve them. By design, these properties are instead checked by the settlement layer post-fill. A contract's failure to enforce one of these properties locally is therefore not a defect; it reflects the intended division of responsibility between on-chain enforcement and off-chain settlement. We call this second class a \emph{settlement exposure}.

\begin{table*}[ht]
\centering
\small
\rowcolors{2}{gray!10}{white}
\begin{adjustbox}{width=\linewidth}
\begin{tabular}{lp{2.1cm}p{4.6cm}p{9cm}}
\toprule
\textbf{ID} & \textbf{Category} & \textbf{Property Enforced} & \textbf{Consequence of Violation} \\
\midrule
AC & Access Control & Fill execution is not gated on a single stored authority (e.g., an owner or admin address). & A compromised or malicious authority can unilaterally block or control fills, undermining the open, permissionless solver role. \\
FRP & Fill Replay Prevention & The same intent cannot be filled twice on the destination chain. & A solver (or attacker) can trigger duplicate payouts against a single locked deposit, draining destination-chain liquidity. \\
DRP & Deposit Replay Prevention & The same intent cannot be deposited or opened twice on the origin chain. & A single locked deposit can back multiple recorded intents, letting an attacker claim solver reimbursement more than once for a single deposit. \\
OCB & Origin Chain Binding & The intent's declared origin chain matches the chain the deposit executes on. & An intent recorded as originating from the wrong chain misleads solvers and the settlement layer about which chain's deposit backs the fill. \\
DCB & Destination Chain Binding & The intent's declared destination chain matches the chain the fill executes on. & A fill can be submitted on a chain the intent never targeted, breaking the correspondence the settlement layer relies on to match fills to deposits. \\
TB & Temporal Binding & A fill cannot execute after the intent's deadline has passed. & An intent can be filled after its deadline, forcing acceptance of terms the depositor no longer agreed to, or filling an intent that should instead be refundable. \\
\bottomrule
\end{tabular}
\end{adjustbox}
\caption{The six invariant violations targeted by \toolname{}.}
\label{tab:invariant-taxonomy}
\end{table*}

Table~\ref{tab:invariant-taxonomy} lists the six invariant violations that this taxonomy identifies as properties an intent-based bridge contract must enforce locally. Section~\ref{sec:evaluation} reports how well \toolname{} detects them in practice. Three further properties are intentionally excluded from this taxonomy: whether the destination-chain recipient, output token, and output amount delivered in a fill actually match what the origin-chain deposit specified. These are settlement exposures, not invariant violations, for the same reason given above: verifying any of the cases above would require access to the origin-chain deposit state at the moment of the fill, which is impossible for a destination-chain contract to do.

\section{System Design}
\label{sec:system-design}

\toolname{} turns the invariant taxonomy of Section~\ref{sec:invariant-taxonomy} into an automated detection pipeline that requires no manual annotation of contracts, and no hand-written test harness of the protocol under analysis. \toolname{} is divided into three phases: Phase 1 statically recovers a protocol's intent structure and identifies the key functions (deposit and fill); Phase 2 performs a static check for centralized signer dependency (AC); and Phase 3 takes Phase 1 output, builds multi-step fuzz sequences, executes them, and checks their outcomes against the defined invariants (FRP, DRP, OCB, DCB, TB). Phase 3 includes a feedback loop that invokes a large language model to recover  protocol-specific input requirements when a deposit or fill call fails unexpectedly. Figure~\ref{fig:architecture} lays out this pipeline end to end.

\usetikzlibrary{positioning, fit, arrows.meta}
\begin{figure*}[t]
\centering
\begin{tikzpicture}[
  node distance=15mm,
  every node/.style={font=\small},
  box/.style={draw, rounded corners, align=center, minimum height=1.1cm,
              text width=3.2cm, inner sep=4pt, fill=white},
  io/.style={draw, rounded corners, align=center, minimum height=0.9cm,
             text width=2.2cm, inner sep=4pt},
  arr/.style={-{Latex[length=2.2mm,width=1.8mm]}, thick},
  phase/.style={draw, dashed, rounded corners, inner sep=7pt}
]

\node[io]  (source)   {*.sol files};
\node[box, right=of source]   (struct)   {Struct \& FieldType Extraction};
\node[box, right=of struct]   (funcrole) {Function Role Classification};
\node[box, right=of funcrole] (a2)       {Centralized Signer Detection (AC)};

\node[box, below=22mm of a2] (fuzzplan) {Fuzz Plan Synthesis and Cross-Step Binding};
\node[box, left=of fuzzplan] (anvil)    {Anvil Execution (Source + Destination Chains)};
\node[box, left=of anvil]    (oracle)   {Invariant Violation Oracle (FRP, DRP, OCB, DCB, TB) };
\node[io,  left=of oracle]   (report)   {Violation Reports};
\node[box,  above=of oracle, below=0.7cm of struct] (llm) {LLM Feedack Loop};

\draw[arr] (source)   -- (struct);
\draw[arr] (struct)   -- (funcrole);
\draw[arr] (funcrole) -- (a2);
\draw[arr] (a2)       -- (fuzzplan);
\draw[arr] (fuzzplan) -- (anvil);
\draw[arr] (anvil)    -- (oracle);
\draw[arr] (oracle)   -- (report);
\draw[arr] (oracle)   -- (llm);
\draw[arr] (llm)   -| (anvil);

\begin{scope}[on background layer]

\node[phase, fit=(struct)(funcrole), fill=blue!10,
      label={above:\textbf{Phase 1: Structure Recovery}}] (p1) {};
\node[phase, fit=(a2), fill=green!10,
      label={above:\textbf{Phase 2: Static Analysis}}] {};
\node[phase, fit=(fuzzplan)(anvil)(oracle)(llm), fill=violet!10,
      label={below:\textbf{Phase 3: Dynamic Analysis}}] {};
\end{scope}

\end{tikzpicture}
\caption{\toolname{}'s three-phase invariant violation detection pipeline.}
\label{fig:architecture}
\end{figure*}
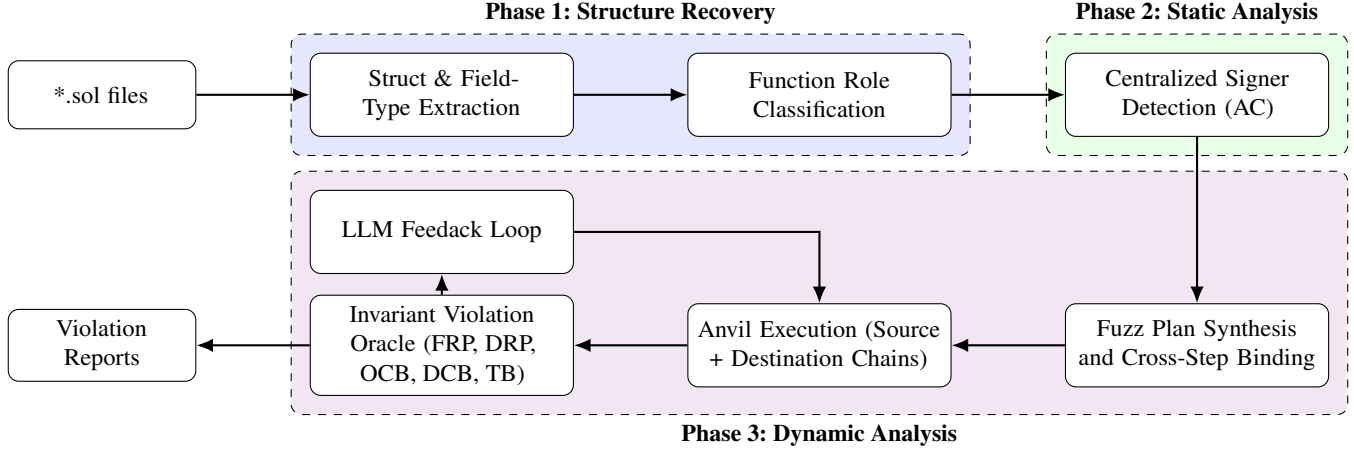

\subsection{Phase 1: Recovering Intent Structure}

\paragraph{Struct and Field Discovery.}

Instead of relying on a human to annotate which struct in a contract represents an intent, and which of its fields carry security-relevant semantics (e.g., an amount, a deadline, a replay-protection identifier), \toolname{} automatically recovers this structure directly from the contract's source code. It uses Slither\cite{feist2019slither}, a static-analysis framework for Solidity with Abstract Syntax Tree (AST) and Intermediate Representation (IR)-level extraction, to extract every struct declared in the contract. Each field is then scored against a set of rules that match name hints and Solidity type hints against a small taxonomy of \texttt{FieldType}s, such as \texttt{UNIQUENESS}, \texttt{TEMPORAL}, \texttt{ORIGIN\_CHAIN}, and \texttt{DEST\_CHAIN}. Each struct's score is the sum of its fields' scores, and the top-scoring candidate struct becomes the contract's canonical intent schema. Hand-crafted scoring is a deliberate choice here, instead of a learning-based model: no labeled corpus of intent structs spanning multiple protocols exists to train one. A rule-based score, moreover, can be inspected field by field to see exactly why a struct was, or was not, selected.

\paragraph{Function Role Classification.}

The second part of Phase 1 is to identify which functions in the codebase are responsible for creating new intents (deposits) and which are responsible for consuming them (fills). This is a nontrivial problem: a protocol may have multiple deposit or fill functions, and each one with different arguments. The key identifier is the functionality of the function, not its name. Therefore, \toolname{} classifies each candidate function by analyzing its Slither IR for a combination of signals. These include whether the function writes a new entry into contract storage (e.g., indicating that it records a fresh intent) and whether it emits events (e.g., a deposit function must emit an event that solvers can observe to know a new intent exists). The primary signal, however, is the direction of the token transfers the function performs: whether it receives or sends funds. A function that transfers funds in, typically alongside creating a new state entry, is classified as a deposit; a function that transfers funds out to an identity-typed recipient is classified as a fill; functions matching neither pattern are left unclassified. All possible pairs of deposit and fill functions are then enumerated, and each pair is passed to Phase 3 for fuzzing. An important note here is the looseness of the heuristics: they are designed to be permissive rather than precise (i.e., prioritizing recall over precision), so that no potential deposit/fill pair is missed. Phase 3's fuzzing and violation detection will later confirm whether a given pair actually behaves as a deposit/fill in practice.

\subsection{Phase 2: Detecting Centralized Signer Dependency (AC)}

Unlike the other five invariants, AC is a static property of the fill function's code rather than a dynamic property of its execution. AC is triggered when the fill function is gated by a single party. Centralization is the leading root cause behind cross-chain bridge exploits historically~\cite{augusto2024sok}, and attacks based on compromised private keys still happen in 2026~\cite{halborn_iotex}.

\toolname{} (Figure~\ref{fig:architecture}) therefore checks AC without executing the contract, using three static detectors. The first two operate directly on Slither's AST and IR. One flags a fill function whose access-control check reduces to an equality comparison against a stored, address-typed state variable, the pattern underlying a check such as comparing the caller against a stored owner or admin address. The other flags a fill function that accepts a signature-shaped byte-string parameter and validates it against a single stored authority reached through the function's internal call chain. The third instead matches on modifier name and argument, since some access-control libraries (e.g., Solady~\cite{solady}) implement a role check as raw storage-slot arithmetic that produces no IR either detector can recognize as a stored-authority comparison. This third detector flags a fill function gated by a modifier restricting execution to an admin/owner address or to a fixed, centrally-managed role set, such as a permissioned solver role.

\subsection{Phase 3: Fuzzing and Violation Detection}

\subsubsection{Fuzz Plan Synthesis}
A hand-written fuzz harness would need a test per invariant per protocol, defeating the goal of a tool that works across arbitrary, previously unseen intent-bridge implementations. \toolname{} (Figure~\ref{fig:architecture}) instead synthesizes a \emph{fuzz plan}: an aggregation of every test sequence generated for a contract, based on a set of templates that target each specific invariant according to the properties present in the intent struct. The number of templates run, therefore, depends on the number of properties/fields identified in Phase 1. For each qualifying deposit/fill pairing identified in Phase 1, the presence of a \texttt{UNIQUENESS}-typed field triggers generation of replay-attempt templates targeting FRP and DRP; the presence of \texttt{ORIGIN\_CHAIN}- or \texttt{DEST\_CHAIN}-typed fields triggers chain-tamper templates targeting OCB and DCB; and the presence of a \texttt{TEMPORAL}-typed field triggers deadline-boundary templates targeting TB, creating and executing intents after the deadline has elapsed. 
Steps are numbered from 0; step 0 is always the template's deposit call. For invariants checked at deposit time alone (OCB and TB's deposit-side variant), this is the step the tamper is applied to directly. For invariants checked later in the sequence (FRP, DRP, DCB, TB's fill-side variant), step 0 is instead an untampered deposit that a later step binds against or replays.

\subsubsection{Cross-Step Parameter Binding}

\paragraph{Binding Relation.}
The easiest way to generate a multi-step call sequence is to generate the parameters freshly in each step independently, but that fails to capture the semantics of a real protocol: a deposit call and the fill call that consumes it must agree on the intent they describe, and a replay-attempt template must resupply the same intent identifier a second time rather than generate a new one. \toolname{} therefore introduces a \emph{binding} mechanism that lets a later step in a template reference a value from an earlier one, either an argument the earlier step's call was invoked with or a value the earlier step's execution generated (e.g., a field of an emitted event). \toolname{} expresses this as a relation between two steps of the same template:
\[
\mathrm{binding}(\mathit{step}_i, \mathit{param}) \;\rightsquigarrow\; \mathrm{src}(\mathit{step}_j, \mathit{arg})
\]
\[
\text{s.t.}\quad j < i, \qquad \mathrm{src} \in \{\mathit{input}, \mathit{event}\}
\]
where $\mathit{param}$ names the call argument $\mathit{step}_i$ must supply, $\mathit{arg}$ names either an argument $\mathit{step}_j$ was invoked with ($\mathit{src}=\mathit{input}$) or a field its execution emitted ($\mathit{src}=\mathit{event}$), and $j < i$ restricts a step to referencing only an earlier one. Here, $\rightsquigarrow$ denotes resolution rather than equality: $\mathit{param}$ takes on whatever value $\mathit{arg}$ held under $\mathit{src}$ once $\mathit{step}_j$ has executed. A parameter with no such binding is instead generated independently by the input-generation tiers of the next subsection and marked \emph{fresh}; this is always true of step 0, and of any later parameter whose correspondence to an earlier step cannot be established, a case detailed below.

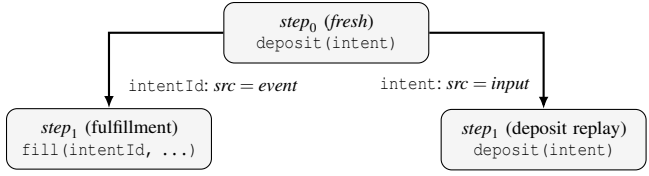
\begin{figure}[t]
\centering
\begin{tikzpicture}[
  node distance=3mm,
  every node/.style={font=\scriptsize},
  box/.style={draw, rounded corners, align=center, text width=2.5cm,
              minimum height=0.8cm, inner sep=3pt, fill=gray!8},
  arr/.style={-{Latex[length=1.6mm,width=1.3mm]}, thick},
  lbl/.style={font=\scriptsize, align=center, text width=2.1cm}
]
\node[box] (s0) {$\mathit{step}_0$ (\emph{fresh})\\ \texttt{deposit(intent)}};
\node[box, below left=0.59cm and 0.15cm of s0] (s1a)
  {$\mathit{step}_1$ (fulfillment)\\ \texttt{fill(intentId, ...)}};
\node[box, below right=0.6cm and 0.15cm of s0] (s1b)
  {$\mathit{step}_1$ (deposit replay)\\ \texttt{deposit(intent)}};

\draw[arr] (s0.west) -| node[lbl, below=25pt, at start]
  {} (s1a.north);
\draw[arr] (s0.east) -| node[lbl, below=17pt, at start]
  {} (s1b.north);

\node[draw=none] at (1.7,-0.69) {\texttt{intent}: $\mathit{src}=\mathit{input}$};
\node[draw=none] at (-1.5,-0.69) {\texttt{intentId}: $\mathit{src}=\mathit{event}$};
\end{tikzpicture}
\caption{Two templates binding against the same $\mathit{step}_0$ deposit. The fulfillment template's \texttt{intentId} is only ever exposed through the deposit's emitted event ($\mathit{src}=\mathit{event}$); the replay template's \texttt{intent} argument must be resupplied byte for byte, which no event carries in full, so it instead binds to $\mathit{step}_0$'s own recorded input ($\mathit{src}=\mathit{input}$).}
\label{fig:binding-example}
\end{figure}

\paragraph{Choosing a Source.}
\toolname{} assigns $\mathit{src}$ once, when the template is built, based on what the earlier function's own event exposes. Deposit and fill functions communicate through events (Figure~\ref{fig:intent-bridge-model}), since an off-chain solver observing a deposit has no access to its calldata, only to what it emitted, so $\mathit{src}=\mathit{event}$ is used whenever the later step's argument corresponds to a logged event field. $\mathit{src}=\mathit{input}$ is the necessary fallback when an event does not carry a value in full: it exposes only the individual fields a contract chooses to log, an identifier, an amount, never a complete struct argument byte for byte, so a later step needing one can only bind to the earlier step's recorded input. Figure~\ref{fig:binding-example} illustrates both sources on a generic deposit/fill pair; unlike the session-intent anchor described below, whose values are independently generated but held consistent across steps, a binding reuses one specific earlier step's value exactly.

\paragraph{Establishing Correspondence.}
Binding is recorded at template-construction time, as a reference to a step, a source, and an argument, before any contract executes. Correspondence is established by static analysis of the earlier function's own source: a later parameter binds to whichever input argument or logged event field whose name most closely matches it, since Solidity's ABI normally preserves parameter and event-field names. Matching prefers an exact normalized-name match, falling back to substring or shared-token overlap, without requiring the two functions to share a naming convention beyond this per-field match. This is not always possible: an ABI-encoded payload passed to a fill call as an opaque byte string, for instance, has no per-field correspondence to the deposit event it is meant to satisfy. When no correspondence can be established, \toolname{} falls back to constructing the argument directly from the session-intent anchor and the schema it encodes, producing a syntactically valid, self-consistent payload rather than leaving the argument unresolved.

\subsubsection{Dual-Layer Tamper Detection}

Leveraging the knowledge on ERC-7683's data structure from Section~\ref{sec:background}, \toolname{} can also detect a subtle class of violations.
Every template described so far treats a \texttt{FieldType} as a single field that is checked at most once, an assumption three ERC-7683 fields violate. \texttt{GaslessCrossChainOrder} contains a nonce, an origin-chain identifier, and an open deadline, and \texttt{OnchainCrossChainOrder} contains a fill deadline. However, these fields can also be repeated in the protocol-specific \texttt{orderData}, which is encoded as a byte string and passed to the deposit function. The contract may check the envelope copy, the inner payload copy, both, or neither. Neither the deadline-boundary template nor the deposit-side deadline template introduced above can distinguish these two copies from a single tampered call: tampering the field once leaves open which of the two, if either,  the contract actually validated. For these three fields specifically, \toolname{} instead generates two variant templates. One tampers only the envelope-level copy, leaving the encoded inner payload untouched; the other decodes the inner payload leveraging \texttt{orderDataType}, tampers only its copy, and re-encodes it, leaving the envelope untouched. Comparing the two outcomes tells the oracle, defined at the end of this subsection, whether the protocol validates the envelope copy, the payload copy, both, or neither, a distinction a single-layer tamper cannot draw. This treatment is specific to ERC-7683's envelope/payload split and to these three fields. A generic deadline field in a protocol with no such split is instead fully covered by the fill-side and deposit-side templates already described.

\subsubsection{Input Generation}\label{subsec:input-generation}
Fields identified by Phase 1 still need concrete values before a template can execute, and neither purely random values nor a single fixed default suffice. A random byte string offered as a struct-typed deposit argument, for instance, never parses as a valid intent, and values chosen independently for a deposit call and a later fill call within the same template can silently describe two different intents rather than one. \toolname{} resolves concrete arguments through a three-tier hierarchy:
\begin{enumerate}
    \item \emph{Session-intent anchor}: a single set of concrete values, generated once per template, from which every step referencing the same schema field draws. This ensures, for example, that a deposit call and the fill call that follows it in the same sequence agree on the amount, the token, and the deadline they describe, rather than each independently sampling its own.
    \item \emph{Boundary-value sampling}: applied whenever no session-intent value applies to a given parameter: values such as zero, one, a maximum safe value, or a chain identifier chosen to match or mismatch the executing fork, selected according to the parameter's \texttt{FieldType}.
    \item \emph{LLM-assisted recovery}: invoked only when a call built from the first two tiers unexpectedly reverts, consistent with prior evidence that LLMs can meaningfully assist with smart-contract security tasks more broadly~\cite{david2023do}. This excludes calls a template deliberately tampers with, and excludes the intentionally-reverting half of a replay or deadline test, since a revert there is the expected, correct outcome. Once triggered, this tier operates as a feedback loop at two scales: within a single recovery attempt, and across the fuzz plan as a whole. Within an attempt, the Solidity revert reason is passed to an LLM (GPT-4o mini, accessed via GitHub Models) behind a common prompt/response interface. The LLM proposes corrected arguments over a bounded number of retries, each attempt informed by the candidate arguments and revert reasons accumulated over earlier retries in the same attempt. A second, coarser-grained loop sits above the first: the result of the first recovery attempt on a given contract function is cached under that function's name, so later templates calling the same function reuse it instead of invoking the LLM again (i.e., a single successful repair benefit every template built on that function).
\end{enumerate}

The third tier is best understood as a cold-start fallback for protocols whose input conventions, a nonstandard fee, an unusual encoding, fall outside what the first two tiers were built to anticipate. The LLM's role is confined to repairing one specific failed call, not to rank templates or decide whether a violation occurred. Section~\ref{sec:evaluation} reports the contributions of this tier.

\subsubsection{Execution and Oracles}

Each template ultimately becomes a sequence of transactions submitted to Anvil, a local Ethereum node, running either a fresh deployment compiled directly from the target contract's source or an existing on-chain deployment reached directly through its ABI and exercised against forked chain state. A template-type-specific oracle then determines whether the sequence's outcome constitutes a violation. A replay-attempt template checks whether a second call carrying the same intent identifier succeeded where FRP or DRP requires it to have reverted. A chain-tamper template checks whether a call carrying a deliberately mismatched chain identifier succeeded where OCB or DCB requires a revert. Either deadline template, fill-side or deposit-side, checks whether the tampered call succeeded where TB requires a revert. The replay oracle in particular has to guard against a protocol-level confound: some protocols assign each deposit an auto-incrementing identifier rather than accepting one the caller supplies. A naive check would therefore flag two structurally identical deposit calls as a replay, even though they produced two distinct, non-colliding intents. \toolname{}'s oracle instead inspects the identifier each call actually produced and reports a violation only when both calls in the same replay template resolve to the \emph{same} identifier. This is the one condition under which a repeated call is a genuine replay rather than an artifact of how the protocol assigns identifiers.

The chain-tamper and deadline oracles face a different confound: a tampered value is only evidence of a missing check if the call actually carries that value into the executed transaction. A target field may have no counterpart on the function under test at all, present only on a struct or an encoded payload used elsewhere in the protocol. In that case, the tamper is well-formed but has nowhere to attach, and the resulting call is indistinguishable from an untampered one. Reporting a violation from such a call would be a false positive: the check was never exercised, let alone found missing. \toolname{} therefore verifies, for every tampered step, that some parameter of the invoked call actually carries the targeted field, directly, as part of a struct argument, or as part of an encoded payload, before allowing its oracle to report a violation. A tamper that cannot attach anywhere is instead recorded as inconclusive, preventing a well-formed but inert tamper on a deployed contract from being misreported as a passed check when the targeted field simply has no counterpart in that contract's call.
\section{Evaluation}
\label{sec:evaluation}

This section evaluates \toolname{}'s ability to recover a protocol's intent structure (Phase 1) and detect the invariant violations of Table~\ref{tab:invariant-taxonomy} (Phases 2--3). We first evaluate \toolname{} against synthetic ground truth with known struct, function roles, and planted bugs. We evaluate it against deployed contracts, where ground truth comes from manual review.

\subsection{Corpus Construction}
\label{sec:eval-corpus}

\begin{figure}[t]
\centering
    \resizebox{0.45\textwidth}{!}{%
      \begin{tikzpicture}[
        node distance=7mm,
        box/.style={draw, rounded corners, align=center, text width=120mm, minimum height=10mm, font=\large},
        arrow/.style={-Triangle, thick}
      ]
      \node[box, fill=blue!10] (start) {\textbf{Total Crawled Repositories} (2,237 contracts in 277 repos)};
      \node[left=of start] (myimage) {\includegraphics[width=0.09\columnwidth]{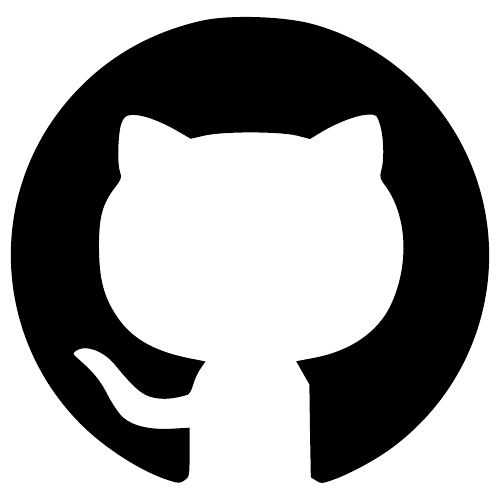}};
      \node[box, below=of start, fill=red!10] (filter) {\textbf{Filter Irrelevant Directories / Files }(864 contracts in 145 repos)};
      \node[left=of filter] (filter_img1) {\includegraphics[width=0.09\columnwidth]{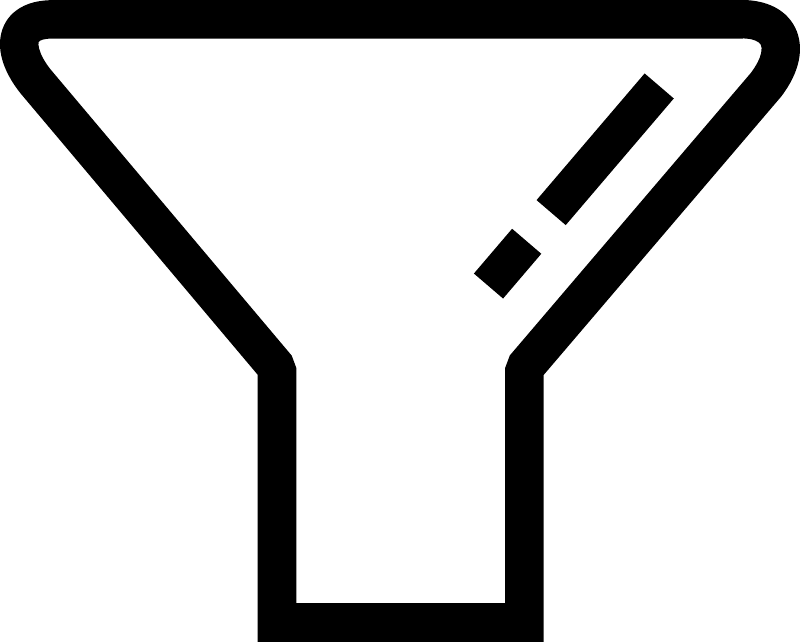}};
      \node[box, below=of filter, fill=red!10] (kw) {\bm{$\ge$} \textbf{3 Keyword Match Filter}  (179 contracts in 69 repos)};
      \node[left=of kw] (filter_img2) {\includegraphics[width=0.09\columnwidth]{figures/filter-outline-icon.pdf}};

      \node[box, below=of kw, fill=red!10] (pre) {\textbf{Standalone-contract filter} (102 contracts in 65 repos)};
    \node[left=of pre] (filter_img3) {\includegraphics[width=0.09\columnwidth]{figures/filter-outline-icon.pdf}};

      \node[box, below=of pre, fill=green!20] (comp) {\textbf{Compile Check} (77 contracts in 51 repos)};
      \node[left=of comp] (filter_img3) {\includegraphics[width=0.09\columnwidth]{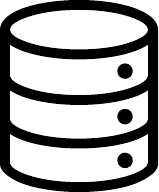}};

      \draw[arrow] (start) -- (filter);
      \draw[arrow] (filter) -- (kw) ;
      \draw[arrow] (kw) -- (pre);
      \draw[arrow] (pre) -- (comp);
      \end{tikzpicture}
    }

\caption{Corpus construction pipeline and filtering stages for the evaluation corpus.}
\label{fig:corpus-pipeline}
\end{figure}

In this work, we assembled a corpus of intent-bridge Solidity contracts from GitHub and filtered down in stages as shown in Figure~\ref{fig:corpus-pipeline}. A crawler first collects candidates via GitHub's code-search API using seventeen intent-bridge-specific Solidity keyword patterns. Because code search matches any file containing these terms anywhere in a repository, a path-based filter removes test, mock, interface, library, and build-artifact files that surface only through incidental keyword matches. A stricter keyword threshold, requiring at least one file to match more than two patterns, then separates repositories with genuine intent-bridge content from those with only incidental overlap, such as token variants, AMMs, and governance tooling that share vocabulary without implementing the pattern. Before Phase 1 analysis, a standalone-contract filter removes files that are abstract contracts, thin wrappers, or chain-specific inheritance extensions of another in-repo contract. Of the 102 contracts left, 25 failed to compile for reasons external to \toolname{} (e.g., missing upstream package dependencies, broken repository build configurations), and are excluded from evaluation. The remaining 77 contracts were manually assigned ground-truth labels (a bridge/not-bridge verdict, the correct intent struct where applicable, and the correct deposit and fill function sets) against which the following subsection reports Phase 1's corpus-scale accuracy.

\subsection{Coverage Measurement}
\label{sec:coverage}
Generic code coverage does not fit invariant testing. Instruction coverage marks a \texttt{JUMPI} covered on mere execution regardless of outcome, while branch coverage requires both outcomes to be observed; but for a security check, the failing branch \emph{is} the violation, so requiring both outcomes ties full branch coverage to the presence of a bug. \toolname{} instead reports \emph{guard-JUMPI coverage}: the fraction of \texttt{JUMPI} sites guarding a \texttt{REVERT} that were visited, regardless of outcome. Guards are identified directly on deployed bytecode. A breadth-first traversal of the control-flow graph, with an abstract-stack interpreter tracking \texttt{PUSH}/\texttt{DUP}/\texttt{SWAP}/\texttt{POP} to resolve jump targets beyond literal pushes, flags any \texttt{JUMPI} whose taken branch or direct fall-through reaches \texttt{REVERT}; this is the pattern \texttt{solc} emits for non-payable and argument-validation checks. Visitation, from an EVM step trace, is computed per entry point (selector resolved via \texttt{evmole}~\cite{evmole}) scoped to only its reachable guards, with proxies resolved first. Section~\ref{sec:eval-llm} reports these figures alongside violation counts.

\subsection{Phase 1: Struct Identification and Function Classification}
\label{sec:eval-phase1}

First, we describe the evaluation setup and then report Phase 1's evaluation results.

\subsubsection{Evaluation Setup}
Phase 1's struct-selection and function-classification accuracy is measured at two scales: (1) a nine-protocol benchmark produced by the authors to serve as a labelled baseline with fully known ground truth, and (2) the corpus described in Section~\ref{sec:eval-corpus}, where ground truth was instead established through manual review. The benchmark comprises five real-world intent bridges (deBridge DLN~\cite{repo_debridge}, Across SpokePool~\cite{repo_across}, Across ERC-7683, Omni SolverNet~\cite{repo_omni}, Relay Depository~\cite{repo_relay}) and four synthetic mock baselines implementing the deposit/fill pattern directly (native-ETH and ERC-20 variants, with and without ERC-7683 envelope encoding), giving every classification decision an unambiguous ground truth.

\subsubsection{Results}

Table~\ref{tab:phase1-benchmark} reports the results on the nine-protocol benchmark: Phase 1 selects the correct intent struct in all nine protocols and achieves 100\% recall on both deposit- and fill-function classification, at 82\% combined precision (33 true positives, 7 false positives, 0 false negatives).

\begin{table}[ht]
\centering
\small
\setlength{\tabcolsep}{6.5pt}
\rowcolors{2}{gray!10}{white}
\begin{adjustbox}{width=\columnwidth}
\begin{tabular}{lrrrrrr}
\toprule
\textbf{Task} & \textbf{TP} & \textbf{FP} & \textbf{FN} & \textbf{Precision} & \textbf{Recall} & \textbf{F1} \\
\midrule
Struct             & 9           & --        & --        & 100\%          & 100\%          & 100\%          \\
Deposit            & 18          & 5         & 0         & 78\%           & 100\%          & 88\%           \\
Fill               & 15          & 2         & 0         & 88\%           & 100\%          & 94\%           \\
\midrule
\rowcolor{white}
\textbf{Combined}  & \textbf{33} & \textbf{7} & \textbf{0} & \textbf{82\%} & \textbf{100\%} & \textbf{90\%} \\
\bottomrule
\end{tabular}
\end{adjustbox}
\caption{Phase 1 accuracy on the nine-protocol benchmark. Struct
selects exactly one candidate per protocol, so no
false-positive/false-negative decomposition applies.}
\label{tab:phase1-benchmark}
\end{table}

Table~\ref{tab:phase1-corpus} reports the same accuracy measures over the 77 manually labeled corpus contracts described in Section~\ref{sec:eval-corpus}. If Phase 1 misses to classify a contract as a bridge (a bridge false negative) it yields no struct or function classification for Phase 1 to be scored on. On the other hand, a contract misclassified as a bridge in Phase 1 (a bridge false positive), has no ground-truth struct or function set to score its output against. For this reason, struct-selection, deposit, and fill accuracy are computed only over the true-positive bridges, the 35 contracts on which Phase 1's verdict and the ground-truth verdict agree that the contract is a bridge. Of these 35 contracts, all carry a struct ground-truth annotation added by the authors, 31 carry a deposit annotation, and 30 carry a fill annotation. Because deposit and fill accuracy are themselves scored at the function level, and a single contract may contribute more than one ground-truth deposit or fill function, the per-task sample sizes are not identical. At corpus scale, Phase 1 again achieves 100\% recall on both deposit- and fill-function classification.

\begin{table}[ht]
\centering
\scriptsize
\setlength{\tabcolsep}{5.5pt}
\rowcolors{2}{gray!10}{white}
\begin{adjustbox}{width=\columnwidth}
\begin{tabular}{lrrrrrrr}
\toprule
\textbf{Task} & \textbf{TP} & \textbf{FP} & \textbf{FN} & \textbf{TN} & \textbf{Precision} & \textbf{Recall} & \textbf{F1} \\
\midrule
Bridge classification & 35 & 9  & 1 & 32 & 79.5\% & 97.2\% & 87.5\% \\
Struct selection       & 31 & 4  & 4 & --  & 88.6\% & 88.6\% & 88.6\% \\
Deposit                & 55 & 21 & 0 & --  & 72.4\% & 100\%  & 84.0\% \\
Fill                   & 31 & 14 & 0 & --  & 68.9\% & 100\%  & 81.6\% \\
\bottomrule
\end{tabular}
\end{adjustbox}
\caption{Phase 1 accuracy over the 77 manually labeled corpus
contracts. TN is defined only for bridge classification, the only
task with a genuine negative class.}
\label{tab:phase1-corpus}
\end{table}

The two false-positive populations differ in composition. At corpus scale, bridge-classification false positives are predominantly adapter or forwarder contracts that call out to an external, separately deployed bridge contract without owning any on-chain intent state themselves. This shape is indistinguishable from a genuine deposit under the criteria Phase 1 uses to identify one, even though the contract is not a bridge in its own right. Deposit- and fill-function false positives within correctly classified bridges are predominantly cross-chain message-handler callbacks: functions a cross-chain messaging protocol invokes to deliver an incoming message. These write state, move funds, and emit events much like a genuine deposit or fill, but they are related to the settlement-layer, not user- or solver-facing entry point. A smaller residual (e.g., auction/bid functions) does not fit either pattern.

Please note that \toolname{} prioritizes recall over precision on deposit- and fill-function classification by design. If Phase 1 fails to classify a function, it never enters a fuzz plan, so no invariant test run for that function. A falsely classified function, in contrast, only costs Phase 3 a wasted fuzzing sequence, which its execution oracle filters out. 

\begin{findingbox}
\textbf{Main Insights.} \toolname{} recovers the correct intent struct and every genuine deposit/fill function with 100\% recall at both benchmark and corpus scale. Its remaining false positives cluster in two explainable patterns, adapter/forwarder contracts and cross-chain message-handler callbacks.
\end{findingbox}

\subsection{Phase 2: Centralized Signer Dependency Detection}
\label{sec:eval-phase2}

In Phase 2, \toolname{} detects whether a fill function is gated on a single stored authority, which poses a centralization risk. This section first describes the evaluation setup and then reports the results.

\subsubsection{Evaluation Setup}

\toolname{} first validates AC detection against a small set of synthetic contracts constructed for this purpose: three variants gating fill execution behind a centralized authority (a stored owner address, a stored verifier address, and an ERC-7683 variant of the same pattern) and four secure baselines with no such gate. Phase 2 is evaluated on a broader real-world corpus: since it requires no forked chain state, it was run over 35 protocol/chain registrations drawn from this work's real-world corpus. This resolves to 21 unique underlying contract sources once the same source analyzed across multiple chain deployments is counted once.

\subsubsection{Results}

\toolname{} flags all three vulnerable variants and none of the four baselines (3/3 recall, 4/4 true negatives). On the crawled corpus, \toolname{} flags 4 of the 21 contracts with an AC finding, each manually verified against the deployed source before being reported here. The remaining 17 are clean. Table~\ref{tab:phase2-corpus} lists the four flagged contracts together with the gate each detector matched: a stored-authority signature check, a direct stored-address equality comparison, and a permissioned-role modifier.

\begin{table}[ht]
\centering
\scriptsize
\rowcolors{2}{gray!10}{white}
\begin{adjustbox}{width=\columnwidth}
\begin{tabular}{p{2.2cm}p{2.8cm}p{2.3cm}}
\toprule
\textbf{Protocol} & \textbf{Fill-Side Gate} & \textbf{Detection Mechanism} \\
\midrule
Relay Depository (Base) & \texttt{allocator.\newline{}isValidSignatureNow(...)} & Stored-authority signature check (IR) \\
Phathdt/simple-bridge \newline SimpleBridge & \texttt{require(msg.sender\newline{}== owner())} & Stored-address equality check (IR) \\
Omni SolverNet \newline (Ethereum mainnet) & \texttt{onlyRoles(SOLVER)} modifier &
Modifier name/argument inspection \\
omni-network/omni \newline SolverNetOutbox \newline (chain 421614) & \texttt{onlyRoles(SOLVER)} modifier & Modifier name/argument inspection \\
\bottomrule
\end{tabular}
\end{adjustbox}
\caption{AC findings on the real-world corpus. Each finding was
manually verified against the deployed contract's source.}
\label{tab:phase2-corpus}
\end{table}

\begin{findingbox}
\textbf{Main Insights.} \toolname{}'s three static detectors achieve perfect recall on the synthetic gated/ungated contracts (100\%) and, without executing any code, flag 4 of 21 real-world protocols with a manually confirmed centralized-signer dependency.
\end{findingbox}

\subsection{Phase 3: Detection Accuracy on the Synthetic Dataset}
\label{sec:eval-phase3}

Phase 3's invariant-violation recall is measured via mutation testing against the four synthetic mock baselines introduced in Section~\ref{sec:eval-phase1} (native-ETH and ERC-20 token transfer, with and without ERC-7683 envelope encoding). For each baseline, this work derives a set of mutants by removing exactly one invariant-enforcing \texttt{require()} guard per mutant (for example, a check against a replay-protection mapping prior to a fill, or a chain-identifier equality check), leaving the underlying state write intact. Group sizes are asymmetric: the ERC-7683 native-ETH baseline contributes 8 mutants (the only group exhibiting the dual-layer envelope/payload variants described in Section~\ref{sec:system-design}), while the remaining three baselines contribute 5 mutants each, for 23 mutants in total. Each contract is compiled and deployed locally. Table~\ref{tab:phase3-mutants} shows the results.

\toolname{} flags all 23 mutants as invariant violations and reports zero violations when run against the four unmutated baselines, yielding 100\% recall and 100\% precision.


\begin{table}[ht]
\centering
\small
\setlength{\tabcolsep}{4pt} 
\rowcolors{2}{gray!10}{white}
\begin{adjustbox}{width=\columnwidth}
\begin{tabular}{lcc}
\toprule
\textbf{Invariant} & \textbf{Total Mutants} & \textbf{Mutants Found} \\
\midrule
FRP (fill replay)              & 4 & 4 \\
DRP (deposit replay)           & 4 & 4 \\
OCB (origin chain binding)      & 6 & 6 \\
DCB (destination chain binding) & 4 & 4 \\
TB (deadline enforcement)       & 5 & 5 \\
\midrule
\rowcolor{white}
\textbf{Total}                         & \textbf{23} & \textbf{23} \\
\bottomrule
\end{tabular}
\end{adjustbox}
\caption{Planted-mutant recall by invariant.}
\label{tab:phase3-mutants}
\end{table}

\subsection{Phase 3: Real-World Evaluation}
\label{sec:eval-fork}

The evaluation setup for Phase 3 evaluation is described first, followed by representative findings.

\subsubsection{Evaluation Setup}

Real intent-based bridge deployments carry protocol-specific initialization, such as role grants, cross-chain messaging registration, and linked periphery contracts, that a generic local redeployment cannot reconstruct. Rather than compiling and redeploying each of the corpus's 77 contracts locally, as in Section~\ref{sec:eval-phase3}'s synthetic mock evaluation, this evaluation forks live chain state and reaches each contract through its real, already-deployed address and ABI. Addresses come from scanning each repository's own deploy-script artifacts, confirming deployed bytecode on the target chain before use. This yields 24 protocol/chain deployments, each run under two input-generation configurations: a baseline (B) using only \toolname{}'s heuristic tiers, and an LLM-assisted (L) configuration adding the recovery tier of Section~\ref{sec:system-design}.

A live, correctly initialized address alone does not make a contract exercisable: executing a deposit function still requires funding the test account with correct token balances and approvals, and, for some deployments, working around an external dependency, such as a cross-chain messaging system.

\subsubsection{Results}

Under the baseline configuration, \toolname{} executed 464 templates across the 24 deployments and confirmed 17 genuine invariant violations, spanning eight vulnerable GitHub repositories, each finding reproducible against the public, deployed bytecode. 
Below, we present three examples of findings with snippets of code to illustrate the missing checks. After, we show the invariant violation breakdown by category.

\paragraph{TB (Deposit-Side) -- Non-Envelope with Undecided Expiration.} eco/eco-routes'~\cite{repo_eco} Portal contract, on Ethereum mainnet, decodes an incoming order's route deadline as part of its \texttt{OrderData} and passes it straight into intent creation without ever comparing it to the current time:

\begin{lstlisting}
OrderData memory orderData =
  abi.decode(order.orderData, (OrderData));
(*\color{red}// FIX: require(orderData.routeDeadline > block.timestamp, "expired");*)
(bytes32 orderId, ) = _publishAndFund(
  orderData.destination, orderData.route,
  orderData.reward, false, msg.sender
);
\end{lstlisting}

Neither \texttt{publish} nor \texttt{\_fundIntent}, the two calls \texttt{\_publishAndFund} makes internally, reference \texttt{orderData.routeDeadline} at any point; an intent whose route deadline has already elapsed is created and funded exactly like a valid one. No solver can ever fill it, but the deposit is accepted and the funds are locked permanently.

\paragraph{TB (Deposit-Side) -- Envelope-Only with Unvalidated ERC-7683 Envelope Deadline.} BootNodeDev's~\cite{repo_bootnodedev} contract resolves an incoming \texttt{OnchainCrossChainOrder} envelope's fill deadline into its inner order record with no accompanying check:

\begin{lstlisting}
OrderData memory orderData = OrderEncoder.decode(_orderData);
(...)
(*\color{red}// FIX: require(\_fillDeadline > block.timestamp, "expired fillDeadline");*)
orderData.fillDeadline = _fillDeadline;
(...)
orderId = OrderEncoder.id(orderData);
(...)
\end{lstlisting}

\toolname{}'s dual-layer tamper mechanism (Section~\ref{sec:system-design}) isolates this gap to the deposit path: \texttt{\_resolvedOrder} never checks that \texttt{\_fillDeadline} lies in the future before writing it into \texttt{orderData.fillDeadline} and accepting the deposit. After a manual analysis, we check that the fill path (i.e., the other smart contract on the destination chain) correctly rejects a fill once this deadline has passed. However, the presence of that check in the destination chain cannot prevent an already-expired intent from being funded in the origin chain; it can only guarantee that intent is never filled. A deposit with an already-elapsed \texttt{\_fillDeadline} therefore locks its funds permanently. This same gap recurs across several deployments.

\paragraph{DRP -- Deposit Replay via a Missing Status Check.} decentxyz's GhostGateway~\cite{repo_ghostgateway}, on Arbitrum, derives its order identifier deterministically from the deposited order's contents through a function \texttt{\_getOrderId} and writes the resulting record without first checking whether that identifier's status is already set:

\begin{lstlisting}
bytes32 orderId = _getOrderId(orderData);
(*\color{red}// FIX: require(orderStatus[orderId] == OrderStatus.UNFILLED, "already opened");*)
orders[orderId] = orderData;
orderStatus[orderId] = OrderStatus.OPENED;
(...)
IERC20(TypeCasts.bytes32ToAddress(orderData.inputToken)).safeTransferFrom(
    msg.sender, address(this), orderData.amountIn
);
\end{lstlisting}

Two deposit calls carrying identical order contents therefore produce the same \texttt{orderId}, and GhostGateway's \texttt{open} function accepts the second call, silently overwriting the first order's record and re-emitting the opening event.

\begin{table}[t]
\centering
\small
\rowcolors{2}{gray!10}{white}
\setlength{\tabcolsep}{4pt}
\begin{adjustbox}{width=\columnwidth}
\begin{tabular}{@{}p{3.8cm}rrrr@{}}
\toprule
\textbf{Invariant} & \textbf{B} & \textbf{L} & \textbf{Arch. (B)} & \textbf{Arch. (L}) \\
\midrule
DRP (deposit replay)         & 4  & 9  & 2   & 2   \\
TB, deposit-side (non-env.) & 4  & 4  & 10  & 10  \\
TB, deposit-side (env.-only) & 9  & 9  & --- & --- \\
\midrule
\rowcolor{white}
\textbf{Total}                & \textbf{17} & \textbf{22} & \textbf{12} & \textbf{12} \\
\bottomrule
\end{tabular}
\end{adjustbox}
\caption{Confirmed invariant violations by category, baseline (B) vs.\ LLM-assisted (L), including Across SpokePool's architectural count under each configuration (Arch.).}
\label{tab:phase3-viol-summary}
\end{table}

\paragraph{Violation Breakdown.}

Table~\ref{tab:phase3-viol-summary} decomposes the 17 baseline violations by category.

\begin{itemize}
    \item \textbf{DRP (4 Instances).} One each on Relay Depository and GhostGateway (the first finding above), and two on eco/eco-routes' Portal contract, whose \texttt{open} and \texttt{openFor} entry points each independently permit deposit replay.
    \item \textbf{TB -- Deposit-Side with Non-Envelope (4 Instances).} All four recur on eco Portal, through two distinct internal paths each on \texttt{open} and \texttt{openFor} (the second finding above).
    \item \textbf{TB -- Deposit-Side with Envelope-Only (9 Instances).} GhostGateway contributes 1; BootNodeDev's two ERC-7683-based implementations, together with the banr1~\cite{repo_banr1}/Kaushikh76~\cite{repo_kaushikh76} deployments, contribute 2 each (6 in total); and MukulKolpe/ETHGlobalTaipei~\cite{repo_mukulkolpe} and SarveshLimaye/trifecta~\cite{repo_sarveshlimaye} contribute 1 each (2 more).
\end{itemize}

Across SpokePool's 12 architectural instances split into TB's 10 non-envelope instances and DRP's 2 \texttt{unsafeDeposit}-only instances. Manual review of the deployed source confirms each is a documented protocol choice: \texttt{\_depositV3} validates only an upper bound on \texttt{fillDeadline}, and the contract's \texttt{INFINITE\_FILL\_DEADLINE} sentinel, used when off-chain validators query for expired deposits, confirms an already-elapsed deadline at deposit time is an anticipated case. \texttt{unsafeDeposit} similarly accepts a caller-supplied intent identifier by design, flagged as such in its own name and documentation. Guard-JUMPI coverage show that these are exercised code paths: 82.9\% of deposit-side and 91.4\% of fill-side guards are reached on both Across chains. 

\begin{figure}[!t]
\centering
\input{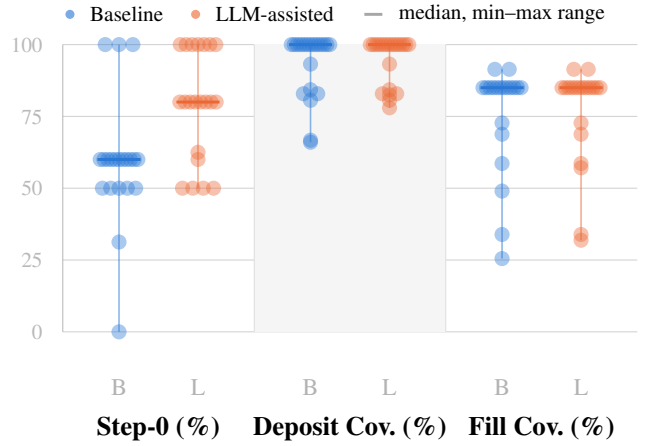}
\caption{Step-0 reachability and guard-JUMPI coverage
(Section~\ref{sec:coverage}), baseline (B) vs.\ LLM-assisted (L).}
\label{fig:phase3-coverage-summary}
\end{figure}

\paragraph{Low Fill-Side Coverage.}

Low fill-side coverage in this evaluation traces to three distinct causes. Of the seven protocol codebases with low or entirely unmeasured fill-side coverage, four, eco Portal and the three dormant corpus deployments below, are low because no call could be constructed or dispatched at all. Two more, Omni SolverNet and Relay Depository, instead reflect a confirmed access-control gate. The seventh, deBridge DLN, is neither: its fill call is constructed and dispatched but fails a fork-specific value check that neither boundary sampling nor LLM correction can supply (detailed below). Figure~\ref{fig:phase3-coverage-summary} plots step-0 reachability and guard-JUMPI coverage under both configurations across the 21 of 24 deployments that completed at least one successful call under either configuration, baseline (B) alongside its LLM-assisted (L) counterpart discussed in Section~\ref{sec:eval-llm}. Each dot is one deployment, jittered to reduce overlap, with the tick marking the median.

\begin{itemize}
    \item \textbf{Confirmed Access-Control Gates.} Every fill call against Omni SolverNet is correctly rejected by the same \texttt{onlyRoles(SOLVER)} gate Section~\ref{sec:eval-phase2} flags under AC (Table~\ref{tab:phase2-corpus}). Coverage there reflects the union of guard checks reached across several differently-shaped rejected calls hitting this gate. This fork evaluation independently corroborates, dynamically, the same centralized-authority dependence Phase 2 already identified statically. Relay Depository's fill-side guards are unreached for the same reason: \texttt{execute()} requires a valid EIP-712 signature from its stored allocator authority, the same AC finding Table~\ref{tab:phase2-corpus} reports, which this evaluation cannot forge.
    \item \textbf{Eco/Eco-Routes' Portal.} Only one of three fill-role candidates (the generic ERC-7683 fill) matches a call the tool can construct, while the two protocol-specific functions, \texttt{fulfill} and \texttt{fulfillAndProve}, genuinely deployed but built around a different, \texttt{Route}-struct-based signature, never dispatch a transaction at all.
    \item \textbf{Dormant Corpus Deployments.} Omni SolverNetOutbox and OpenGateV2~\cite{repo_dimasriat} never reach a fill call at all (1 and 2 templates, respectively): their deposit step reverts. These hackathon-style deployments depend on on-chain protocol state, cross-chain messaging registration among them, that a generic fork cannot reconstruct.
    \item \textbf{deBridge DLN.} \texttt{fulfillOrder} requires the order's \texttt{takeChainId} field to match the executing fork's actual chain identifier, a value boundary-value sampling does not reliably supply; a large share of fill calls revert before reaching any invariant check. LLM-assisted recovery does not close this either, since a fork-specific chain identifier is not something a Solidity revert message can supply.
\end{itemize}

\subsubsection{Performance Metrics}

Static extraction (Phase 1) takes a median 0.47s per contract across 27 mock contracts (23 mutants, 4 baselines; end-to-end evaluation there (Phases 1-3) takes a median 0.88s per contract, over which 273 templates (507 transactions) execute in a median 14ms, with no network round-trip. Phase 3 alone, on real-world crawled contracts over public RPC, takes a median 26.4s per deployment (varies with template count): 1.7s/template and 0.9s/transaction, roughly two orders of magnitude slower, since each transaction costs a very slow public RPC round-trip plus a \texttt{debug\_traceTransaction} call for guard-JUMPI coverage. The smallest genuine finding, Relay Depository's DRP, is confirmed in 3.8s (0.96s/template, 6 transactions) against an already-forked chain; a larger deployment, BootNodeDev's Hyperlane7683, runs 17 templates in 31.16s (1.83s/template, 26 transactions).

\begin{findingbox}
\textbf{Main Insights.} Forking 24 live protocol deployments, \toolname{} confirms 17 genuine invariant violations across 8 vulnerable GitHub repositories, each independently reproducible against public deployed bytecode, while correctly separating Across SpokePool's 12 flagged instances as a documented architectural choice.
\end{findingbox}

\subsection{Contribution of LLM Feedback Loop}
\label{sec:eval-llm}

\paragraph{Overall Impact.}

The LLM-assisted recovery tier enabled on the same 24 deployments raises step-0 success (the fraction of deposit/fill calls that execute successfully rather than reverting before invariants are tested), from 167/307 (54.4\%) to 261/307 (85.0\%), over 81 LLM recovery calls (with a total cost  $\approx$\$0.05\footnote{princing of \$0.15 per million input tokens}), and raises the invariant-violation count from 17 to 22 (cf. Table~\ref{tab:phase3-viol-summary}).

All 5 new violations are DRP (deposit replay), reached through native-currency deposit paths the baseline configuration could not: eco/eco-routes' Portal contract's \texttt{fund}, \texttt{fundFor}, \texttt{publishAndFund}, and \texttt{publishAndFundFor} entry points (4), and Relay Depository's \texttt{depositErc20} entry point (1). The latter mirrors the already-confirmed replay on its \texttt{depositNative} counterpart. Reaching each entry point requires supplying the correct native fee or token-transfer amount the call expects, a value neither the session-intent anchor nor boundary-value sampling can infer from the ABI alone. This tier also lets \texttt{fillRelay} and \texttt{fillV3Relay} succeed on Across SpokePool by correcting its multi-field \texttt{relayData} struct, raising step-0 success on this deployment to 35/35; its architectural total is unchanged at 12.

\paragraph{Coverage Gains and Their Limits.}


deBridge DLN illustrates this tier's contribution: step-0 success rises from 0\% to 63\%, deposit-side coverage rises from 66.0\% to 78.0\% and fill-side coverage from 49.0\% to 57.1\%. eco/eco-routes' Portal shows the same pattern at larger scale: step-0 success rises from 31\% to a fully saturated 100\%, and deposit-side coverage from 66.7\% to 82.5\%.

\paragraph{Deposit- vs.\ Fill-Side Asymmetry.}

This tier's contribution concentrates on protocols whose deposit and fill functions expect input conventions, a native fee, a fork-specific gas or messaging cost, that the first two input-generation tiers cannot get from the intent struct's field types. This substantiates the cold-start framing of Section~\ref{subsec:input-generation}. Its effect on fill-side reachability is narrower: aside from Across and eco/eco-routes just discussed, fill-side guard-JUMPI coverage is unchanged across every other protocol. Section~\ref{sec:discussion} discusses why this asymmetry is expected: a Solidity revert reason is a causal signal that often names the value the failing check expected, which is why this tier succeeds on self-contained deposit-side calls. A fill call in most protocols examined, by contrast, is missing structured values, an \texttt{orderId} or a re-encoded payload, that exist only in the deposit transaction's event log.

\begin{findingbox}
\textbf{Main Insights.} The LLM-assisted tier deepens deposit-side testing, raising step-0 success from 54.4\% to 85.0\% and surfacing five additional genuine DRP violations by inferring protocol-specific fees and amounts directly from transaction revert reasons.
\end{findingbox}
\section{Discussion}
\label{sec:discussion}

\paragraph{Coverage Only as a Measurement Tool.}
A natural question is why the guard-JUMPI coverage measurement already collected (Section~\ref{sec:coverage}) is not fed back to drive input generation directly. The evaluation already tests a strictly more capable version of this idea: the LLM-assisted recovery tier also proposes new arguments for a failed call, guided by the actual revert reason rather than only a covered/uncovered signal, yet Section~\ref{sec:eval-llm} reports its effect on fill-side coverage is unchanged across nearly every protocol family. This is because the residual fill-side gaps are not argument-guessing problems: most fill arguments are already resolved directly through the cross-step binding mechanism of Section~\ref{sec:system-design}, and the rest fail for structural reasons, an access-control gate, on-chain state a generic fork cannot reconstruct, or a fill signature with no counterpart in the recovered schema, none of which a better-guided guess, coverage-driven or otherwise, can supply. A coverage-guided search loop would therefore add search overhead without adding tested surface area.

\paragraph{Fill Functions with Protocol-Specific Argument Shapes.}

\toolname{}'s fill-argument construction targets the parameters a generic ERC-7683 fill call carries, \texttt{fill(orderId, originData, fillerData)}, or their equivalent recovered from the deposit-side intent schema; it cannot construct arguments for a fill entry point whose parameters have no such counterpart. eco/eco-routes' Portal contract illustrates this: alongside the generic fill entry point, it also exposes protocol-specific \texttt{fulfill}/\texttt{fulfillAndProve} functions taking a \texttt{Route} struct, a \texttt{prover} address, and a \texttt{sourceChainDomainID}, none of which corresponds to any field in the recovered intent schema. Closing this gap would mean hardcoding eco-routes' own proof-hash formula (\texttt{keccak256(abi.encodePacked(chainId, keccak256(abi.encode(route)), rewardHash))}) directly into the harness: unlike ERC-7683's \texttt{orderData} encoding, which is standardized and derivable from the recovered schema alone, this construction is bespoke to eco-routes, and hardcoding it is exactly the per-protocol special-casing \toolname{} is designed to avoid. The LLM-assisted recovery tier cannot close this gap either, since it repairs calls by inferring arguments from revert reasons. This check instead requires reproducing a deterministic hash's exact byte-level preimage, and no language model can guess a 32-byte digest into agreement with a value it cannot itself compute.



\section{Related Work}
\label{sec:related-work}

No existing tools detect the invariant violations of Table~\ref{tab:invariant-taxonomy} without a hand-written, protocol-specific oracle, since no tool models the deposit/fill/settlement structure that intent-based bridges assume. Coverage-driven fuzzers, Echidna~\cite{grieco2020echidna}/Medusa~\cite{trailofbits2024medusa} (property-based) and Foundry fuzz~\cite{foundry2024} (unit-test), only flag a violation of a property already encoded as an assertion. Foundry fuzz additionally has no mechanism for the multi-step deposit$\to$fill$\to$replay sequences or ERC-7683's envelope/inner-payload structure these invariants require. Recent work on LLM-guided fuzzing~\cite{jiang2024fuzzing,oliinyk2024fuzzing,eom2024fuzzing} targets general software, JavaScript engines, and embedded firmware rather than smart contracts. Static analyzers such as Slither~\cite{feist2019slither} instead match known-bad code patterns (reentrancy, arbitrary-send-eth, timestamp dependence), i.e., removing a \texttt{require()} guard that enforces a protocol-level invariant leaves no new pattern to match, so a pattern-matching detector cannot, by construction, recognize the resulting \emph{absence} of a check. Even fuzzers built for stronger coverage, such as EFCF~\cite{DBLP:conf/eurosp/RodlerPLBHKD23} (cross-contract interaction coverage) and ConFuzzius~\cite{torres2021confuzzius} (concolic execution), remain confined to a single chain, with no mechanism for cross-chain replay or chain-binding sequences.

A second line of work targets bridge contracts directly rather than generic Solidity, but not intent-based bridges specifically: BridgeGuard~\cite{zhou2025bridgeguard} uses symbolic dataflow analysis to find missing relayer/oracle authentication; SmartAxe~\cite{liao2024smartaxe} infers access-control patterns and builds cross-chain control-/data-flow graphs to catch semantic inconsistencies; BridgeFuzz~\cite{DBLP:conf/eurosec/WinklerSGKD26}, motivated by prior tools' single-contract focus missing the cross-chain interaction chain, targets the hybrid on-chain-contract/off-chain-relayer architecture to detect balance mismatches, protocol errors, and off-chain denial-of-service bugs. All three are built on the original bridge model that intent-based bridges replace. Since none of the solutions above model the deposit/fill/settlement structure central to intent-based bridges, nor the invariants in Table~\ref{tab:invariant-taxonomy}, we do not attempt an empirical comparison.


Finally, work on intent-based bridges is still scarce. Beyond some empirical papers on this topic,~\cite{augusto2026exploiting} has examined the economic exploitation of solver liquidity.
\section{Conclusion}
\label{sec:conclusion}

This paper presented \toolname{}, a protocol-aware fuzzer for intent-based bridges, and four concrete contributions. First, a taxonomy separating invariant violations from settlement exposures (the latter, are delegated to the off-chain settlement layer and are out of scope). Second, \toolname{} recovers a bridge's intent structure and deposit/fill function roles directly from unannotated Solidity source, at 100\% struct-recovery and function-classification recall and 82\% combined precision on a 9-protocol benchmark (5 real-world, 4 synthetic), and 97.2\%/88.6\% bridge-classification and struct-selection recall with 100\% deposit/fill recall at corpus scale (Section~\ref{sec:eval-phase1}). Third, it synthesizes multi-step fuzz sequences from the recovered structure alone, including a cross-step binding mechanism that carries values between steps and a dual-layer tamper mechanism that isolates whether an ERC-7683 envelope, its encoded inner payload, or both are actually validated (Section~\ref{sec:system-design}). Fourth, a three-tier input-generation hierarchy, anchored on a shared session intent, backed by boundary-value sampling, and falling back to an LLM-assisted recovery step. Across synthetic ground truth and 24 production intent-based bridge deployments, \toolname{} finds 22 real invariant violations independently reproducible against publicly deployed bytecode, without any hand-written or protocol-specific properties. These results indicate that automatically recovering intent-based bridge structure and using it to drive fuzzing is both practical and effective at surfacing genuine, exploitable weaknesses in the intent-based bridge ecosystem as it is deployed today.

\cleardoublepage
\cleardoublepage

\bibliographystyle{plainurl}
\bibliography{references}

@inproceedings{grieco2020echidna,
  author    = {Gustavo Grieco and Will Song and Artur Cygan and Josselin Feist and Alex Groce},
  title     = {Echidna: Effective, Usable, and Fast Fuzzing for Smart Contracts},
  booktitle = {Proceedings of the 29th ACM SIGSOFT International Symposium on Software Testing and Analysis (ISSTA)},
  year      = {2020},
  pages     = {557--560},
  publisher = {ACM},
  address   = {Virtual Event, USA},
  doi       = {10.1145/3395363.3404366}
}

@misc{trailofbits2024medusa,
  author       = {{Trail of Bits}},
  title        = {Medusa: Parallelized, Coverage-Guided, Mutational {Solidity} Smart Contract Fuzzing, Powered by go-ethereum},
  howpublished = {\url{https://github.com/crytic/medusa}},
  year         = {2024},
  note         = {Accessed 2026}
}

@misc{foundry2024,
  author       = {{Foundry Contributors}},
  title        = {Foundry: A Blazing Fast, Portable, and Modular Toolkit for {Ethereum} Application Development},
  howpublished = {\url{https://github.com/foundry-rs/foundry}},
  year         = {2024},
  note         = {Accessed 2026}
}

@inproceedings{feist2019slither,
  author    = {Josselin Feist and Gustavo Grieco and Alex Groce},
  title     = {Slither: A Static Analysis Framework for Smart Contracts},
  booktitle = {Proceedings of the 2nd IEEE/ACM International Workshop on Emerging Trends in Software Engineering for Blockchain (WETSEB)},
  year      = {2019},
  pages     = {8--15},
  address   = {Montreal, Canada},
  doi       = {10.1109/WETSEB.2019.00008}
}

@inproceedings{jiang2018contractfuzzer,
  author    = {Bo Jiang and Ye Liu and W. K. Chan},
  title     = {ContractFuzzer: Fuzzing Smart Contracts for Vulnerability Detection},
  booktitle = {Proceedings of the 33rd ACM/IEEE International Conference on Automated Software Engineering (ASE)},
  year      = {2018},
  pages     = {259--269},
  address   = {Montpellier, France},
  publisher = {ACM},
  doi       = {10.1145/3238147.3238177}
}

@inproceedings{torres2021confuzzius,
  author    = {Christof Ferreira Torres and Antonio Ken Iannillo and Arthur Gervais and Radu State},
  title     = {{ConFuzzius}: A Data Dependency-Aware Hybrid Fuzzer for Smart Contracts},
  booktitle = {Proceedings of the 2021 IEEE European Symposium on Security and Privacy (EuroS\&P)},
  year      = {2021},
  pages     = {103--119},
  address   = {Vienna, Austria},
  doi       = {10.1109/EuroSP51992.2021.00018}
}

@ARTICLE{zhou2025bridgeguard,
  author={Zhou, Zequan and Luo, Xiling and Ji, Xiaohai and Mao, Jian and He, Ting and Wang, Junjun and Wu, Qianhong},
  journal={IEEE Transactions on Dependable and Secure Computing},
  title={{BridgeGuard}: Checking External Interaction Vulnerabilities in Cross-Chain Bridge Router Contracts Based on Symbolic Dataflow Analysis},
  year={2025},
  volume={22},
  number={5},
  pages={5798--5812},
  doi={10.1109/TDSC.2025.3576114}
}

@article{liao2024smartaxe,
  author = {Liao, Zeqin and Nan, Yuhong and Liang, Henglong and Hao, Sicheng and Zhai, Juan and Wu, Jiajing and Zheng, Zibin},
  title = {{SmartAxe}: Detecting Cross-Chain Vulnerabilities in Bridge Smart Contracts via Fine-Grained Static Analysis},
  year = {2024},
  issue_date = {July 2024},
  publisher = {Association for Computing Machinery},
  address = {New York, NY, USA},
  volume = {1},
  number = {FSE},
  url = {https://doi.org/10.1145/3643738},
  doi = {10.1145/3643738},
  journal = {Proc. ACM Softw. Eng.},
  month = jul,
  articleno = {12},
  numpages = {22}
}

@inproceedings{augusto2024sok,
  author    = {Andr\'{e} Augusto and Rafael Belchior and Miguel Correia and Andr\'{e} Vasconcelos and Luyao Zhang and Thomas Hardjono},
  title     = {{SoK}: Security and Privacy of Blockchain Interoperability},
  booktitle = {Proceedings of the 45th IEEE Symposium on Security and Privacy (S\&P)},
  year      = {2024}
}

@article{augusto2026exploiting,
  title={Exploiting liquidity exhaustion attacks in intent-based cross-chain bridges},
  author={Augusto, Andr{\'e} and Torres, Christof Ferreira and Vasconcelos, Andr{\'e} and Correia, Miguel},
  journal={arXiv preprint arXiv:2602.17805},
  year={2026}
}

@inproceedings{lee2023sok,
  author    = {Sung-Shine Lee and Alexandr Murashkin and Martin Derka and Jan Gorzny},
  title     = {{SoK}: Not Quite Water Under the Bridge: Review of Cross-Chain Bridge Hacks},
  booktitle = {Proceedings of the 2023 IEEE International Conference on Blockchain and Cryptocurrency (ICBC)},
  year      = {2023},
  address   = {Dubai, UAE},
  doi       = {10.1109/ICBC56567.2023.10174993}
}

@inproceedings{zhang2024sok,
  author    = {Mengya Zhang and Xiaokuan Zhang and Josh Barbee and Yinqian Zhang and Zhiqiang Lin},
  title     = {{SoK}: Security of Cross-Chain Bridges: Attack Surfaces, Defenses, and Open Problems},
  booktitle = {Proceedings of the 27th International Symposium on Research in Attacks, Intrusions and Defenses (RAID)},
  year      = {2024},
  publisher = {ACM},
  doi       = {10.1145/3678890.3678894},
  note      = {arXiv:2312.12573}
}

@misc{erc7683,
  author       = {Francisco Giordano and Mark Toda and Matt Rice and Nick Pai and Alexander Lindgren and Mark Gretzke and Chris Cashwell},
  title        = {{ERC-7683}: Cross Chain Intents},
  howpublished = {Ethereum Improvement Proposals, no. 7683},
  year         = {2024},
  month        = apr,
  url          = {https://eips.ethereum.org/EIPS/eip-7683},
  note         = {Draft}
}

@inproceedings{wu2024we,
  title={Are we there yet? unraveling the state-of-the-art smart contract fuzzers},
  author={Wu, Shuohan and Li, Zihao and Yan, Luyi and Chen, Weimin and Jiang, Muhui and Wang, Chenxu and Luo, Xiapu and Zhou, Hao},
  booktitle={Proceedings of the IEEE/ACM 46th international conference on software engineering},
  pages={1--13},
  year={2024}
}

@misc{repo_across,
  author       = {{Across Protocol}},
  title        = {across-protocol/contracts},
  howpublished = {\url{https://github.com/across-protocol/contracts}},
  year         = {2026},
  note         = {GitHub repository, accessed 2026}
}

@misc{repo_debridge,
  author       = {{deBridge Finance}},
  title        = {debridge-finance/dln-contracts},
  howpublished = {\url{https://github.com/debridge-finance/dln-contracts}},
  year         = {2026},
  note         = {GitHub repository, accessed 2026}
}

@misc{repo_omni,
  author       = {{Omni Network}},
  title        = {omni-network/omni},
  howpublished = {\url{https://github.com/omni-network/omni}},
  year         = {2026},
  note         = {GitHub repository, accessed 2026}
}

@misc{repo_relay,
  author       = {{Relay Protocol}},
  title        = {relayprotocol/relay-depository},
  howpublished = {\url{https://github.com/relayprotocol/relay-depository}},
  year         = {2026},
  note         = {GitHub repository, accessed 2026}
}

@misc{repo_eco,
  author       = {{eco}},
  title        = {eco/eco-routes},
  howpublished = {\url{https://github.com/eco/eco-routes}},
  year         = {2026},
  note         = {GitHub repository, accessed 2026}
}

@misc{repo_ghostgateway,
  author       = {{decentxyz}},
  title        = {decentxyz/decentents},
  howpublished = {\url{https://github.com/decentxyz/decentents}},
  year         = {2026},
  note         = {GitHub repository, accessed 2026}
}

@misc{repo_bootnodedev,
  author       = {{BootNodeDev}},
  title        = {BootNodeDev/intents-framework},
  howpublished = {\url{https://github.com/BootNodeDev/intents-framework}},
  year         = {2026},
  note         = {GitHub repository, accessed 2026}
}

@misc{repo_banr1,
  author       = {{banr1}},
  title        = {banr1/forked-intents-framework},
  howpublished = {\url{https://github.com/banr1/forked-intents-framework}},
  year         = {2026},
  note         = {GitHub repository, accessed 2026}
}

@misc{repo_kaushikh76,
  author       = {{Kaushikh76}},
  title        = {Kaushikh76/intents-framework},
  howpublished = {\url{https://github.com/Kaushikh76/intents-framework}},
  year         = {2026},
  note         = {GitHub repository, accessed 2026}
}

@misc{repo_mukulkolpe,
  author       = {{MukulKolpe}},
  title        = {MukulKolpe/ETHGlobalTaipei},
  howpublished = {\url{https://github.com/MukulKolpe/ETHGlobalTaipei}},
  year         = {2026},
  note         = {GitHub repository, accessed 2026}
}

@misc{repo_sarveshlimaye,
  author       = {{SarveshLimaye}},
  title        = {SarveshLimaye/trifecta},
  howpublished = {\url{https://github.com/SarveshLimaye/trifecta}},
  year         = {2026},
  note         = {GitHub repository, accessed 2026}
}

@misc{repo_dimasriat,
  author       = {{dimasriat}},
  title        = {dimasriat/fuelstack},
  howpublished = {\url{https://github.com/dimasriat/fuelstack}},
  year         = {2026},
  note         = {GitHub repository, accessed 2026}
}

@misc{lifi_intents,
  title={LI.FI - With Intents, It’s Solvers All The Way Down},
  author={Arjun Chand},
  url={https://li.fi/knowledge-hub/with-intents-its-solvers-all-the-way-down},
  abstractNote={null},
  journal={LI.FI},
  language={en}
}

@misc{evmole,
  author       = {{Maxim Andreev}},
  title        = {cdump/evmole},
  howpublished = {\url{https://github.com/cdump/evmole}},
  year         = {2026},
  note         = {GitHub repository, accessed 2026}
}

@misc{solady,
  author       = {Solady},
  title        = {vectorized/solady},
  howpublished = {\url{https://github.com/vectorized/solady}},
  year         = {2026},
  note         = {GitHub repository, accessed 2026}
}

@misc{halborn_iotex,
  author       = {Halborn},
  title        = {Explained: The IoTeX Hack (February 2026)},
  howpublished = {\url{https://www.halborn.com/blog/post/explained-the-iotex-hack-february-2026}},
  year         = {2026},
}

@article{Miller:90,
  author  = {Barton P. Miller and Lars Fredriksen and Bryan So},
  title   = {An Empirical Study of the Reliability of {UNIX} Utilities},
  journal = {Communications of the ACM},
  volume  = {33},
  number  = {12},
  year    = {1990},
  pages   = {32--44},
  publisher = {ACM},
  doi     = {10.1145/96267.96279}
}

@inproceedings{jiang2024fuzzing,
  title     = {When fuzzing meets {LLMs}: Challenges and opportunities},
  author    = {Jiang, Y. and Liang, J. and Ma, F. and Chen, Y. and Zhou, C. and Shen, Y. and Wu, Z. and Fu, J. and Wang, M. and Li, S. and Zhang, Q.},
  booktitle = {Companion Proceedings of the 32nd ACM International Conference on the Foundations of Software Engineering},
  pages     = {492--496},
  year      = {2024}
}

@inproceedings{oliinyk2024fuzzing,
  title     = {Fuzzing {BusyBox}: Leveraging {LLM} and crash reuse for embedded bug unearthing},
  author    = {Oliinyk, Y. and Scott, M. and Tsang, R. and Fang, C. and Homayoun, H.},
  booktitle = {33rd USENIX Security Symposium (USENIX Security 24)},
  pages     = {883--900},
  year      = {2024}
}

@inproceedings{eom2024fuzzing,
  title     = {Fuzzing javascript interpreters with coverage-guided reinforcement learning for {LLM}-based mutation},
  author    = {Eom, J. and Jeong, S. and Kwon, T.},
  booktitle = {Proceedings of the 33rd ACM SIGSOFT International Symposium on Software Testing and Analysis},
  pages     = {1656--1668},
  year      = {2024}
}

@article{david2023do,
  author = {I. David and L. Zhou and K. Qin and D. Song and L. Cavallaro and A. Gervais},
  title  = {Do you still need a manual smart contract audit?},
  month  = jun,
  number = {arXiv:2306.12338v2},
  year   = {2023}
}

@article{chitra2024analysis,
  title={An analysis of intent-based markets},
  author={Chitra, Tarun and Kulkarni, Kshitij and Pai, Mallesh and Diamandis, Theo},
  journal={arXiv preprint arXiv:2403.02525},
  year={2024}
}

@inproceedings{DBLP:conf/eurosp/RodlerPLBHKD23,
  author       = {Michael Rodler and
                  David Paa{\ss}en and
                  Wenting Li and
                  Lukas Bernhard and
                  Thorsten Holz and
                  Ghassan Karame and
                  Lucas Davi},
  title        = {EFCF: High Performance Smart Contract Fuzzing for
                  Exploit Generation},
  booktitle    = {8th {IEEE} European Symposium on Security and Privacy, EuroS{\&}P
                  2023, Delft, Netherlands, July 3-7, 2023},
  pages        = {449--471},
  publisher    = {{IEEE}},
  year         = {2023},
  url          = {https://doi.org/10.1109/EuroSP57164.2023.00034},
  doi          = {10.1109/EUROSP57164.2023.00034},
  bibsource    = {dblp computer science bibliography, https://dblp.org}
}

@inproceedings{DBLP:conf/eurosec/WinklerSGKD26,
  author       = {Pascal Winkler and
                  Christian Scholz and
                  Jens{-}Rene Giesen and
                  Noah Kappert and
                  Lucas Davi},
  title        = {Fuzzing Cross-Chain Vulnerabilities with BridgeFuzz},
  booktitle    = {Proceedings of the 19th European Workshop on Systems Security, EuroSec
                  2026, Edinburgh, Scotland, UK, April 27-30, 2026},
  pages        = {81--88},
  publisher    = {{ACM}},
  year         = {2026},
  url          = {https://doi.org/10.1145/3803525.3804980},
  doi          = {10.1145/3803525.3804980},
  bibsource    = {dblp computer science bibliography, https://dblp.org}
}

@InProceedings{10.1007/978-3-032-32575-4_16,
    author="Winkler, Pascal and Giesen, Jens-Rene and Draissi, Oussama and Badaloni, Federico and Holler, Sebastian and Schneidewind, Clara and Davi, Lucas",
    editor="Jee, Kangkook
    and Kate, Aniket",
    title="{\$}2B Lessons: Brigade as a Defense Against Real-World DeFi Bridge Exploits",
    booktitle="Applied Cryptography and Network Security",
    year="2027",
    publisher="Springer Nature Switzerland",
    address="Cham",
    pages="423--452",
    isbn="978-3-032-32575-4"
}

\end{document}